# 20-80nm Channel Length InGaAs Gate-all-around Nanowire MOSFETs with EOT=1.2nm and Lowest SS=63mV/dec

J. J. Gu,[1] X. W. Wang,[2] H. Wu,[1] J. Shao,[3] A. T. Neal,[1] M. J. Manfra,[3] R. G. Gordon,[2] and P. D. Ye [1]

[1] School of Electrical and Computer Engineering and Birck Nanotechnology Center, Purdue University, West Lafayette, IN 47906, U.S.A.
[2] Department of Chemistry and Chemical Biology, Harvard University, Cambridge, MA 02138, U.S.A.
[3] Department of Physics, Purdue University, West Lafayette, IN, 47906, U.S.A.
Tel: 1-765-494-7611, Fax: 1-765-496-7443, Email: yep@purdue.edu, jjgu@purdue.edu

## Abstract

In this paper, 20nm - 80nm channel length ($L_{ch}$) InGaAs gate-all-around (GAA) nanowire MOSFETs with record high on-state and off-state performance have been demonstrated by equivalent oxide thickness (EOT) and nanowire width ($W_{NW}$) scaling down to 1.2nm and 20nm, respectively. SS and DIBL as low as 63mV/dec and 7mV/V have been demonstrated, indicating excellent interface quality and scalability. Highest $I_{ON}$ = 0.63mA/μm and $g_m$ = 1.74mS/μm have also been achieved at $V_{DD}$=0.5V, showing great promise of InGaAs GAA technology for 10nm and beyond high-speed low-power logic applications.

## Introduction

Recently, a top-down technology for III-V GAA nanowire MOSFETs has been demonstrated [1-2]. However, the device metrics such as $g_m$, $V_{DD}$, SS, DIBL, and the $L_{ch}$ scaling of the III-V GAA devices demonstrated in [1] are still greatly limited by the large EOT of the devices. In this paper, we experimentally demonstrate InGaAs GAA nanowire MOSFETs with an EOT down to 1.2nm by the successful integration of ternary oxide dielectric LaAlO$_3$ (k~16) [3]. The reduction of EOT has allowed the demonstration of the first 20nm $L_{ch}$ InGaAs MOSFETs with $g_m$ as high as 1.74mS/μm at $V_{ds}$=0.5V and negligible short channel effects (SCE). A systematic scaling metrics study with $L_{ch}$ between 20-80nm and nanowire size-dependent transport study with $W_{NW}$ between 20-35nm has also been carried out for three different gate stacks, demonstrating near-ideal SS=63mV/dec and DIBL=7mV/V. It is shown that the integration of 4nm LaAlO$_3$ with ultra-thin 0.5nm Al$_2$O$_3$ interfacial layer allow reduction of EOT to 1.2nm with optimized interface trap density ($D_{it}$), offering excellent scalability, near-ballistic transport, and high $g_m$ at low supply voltage.

## Experiment

Fig. 1 shows a diagram of an InGaAs GAA nanowire MOSFET fabricated in this work. The starting material is a 30nm lightly p-doped InGaAs channel layer, an 80nm undoped InP sacrificial layer, and a 100nm undoped InAlAs etch stop layer on semi-insulating InP (100) substrate grown by molecular beam epitaxy (MBE). The InGaAs channel layer consists of 10nm In$_{0.53}$Ga$_{0.47}$As layer sandwiched by two 10nm In$_{0.65}$Ga$_{0.35}$As layers to boost the channel mobility and reduce the $D_{it}$. Key device dimensions are shown in Fig. 1 and the process flow is shown in Fig. 2. The three gate stacks studied are highlighted in Figs. 1-2. Samples A and B have a 0.5nm Al$_2$O$_3$/4nm LaAlO$_3$ stack (EOT = 1.2nm), where Al$_2$O$_3$ was grown before LaAlO$_3$ for sample A and vice versa for sample B in order to study the effect of ultra-thin Al$_2$O$_3$ passivation layer on the LaAlO$_3$/InGaAs interface quality. Sample C has 3.5nm Al$_2$O$_3$ gate (EOT = 1.7nm). The fabrication process is similar to that demonstrated in [1]. The additional InAlAs bottom layer provides better control of the selective wet etching in the nanowire release process and cuts off parasitic leakage through the substrate. The Al$_2$O$_3$/LaAlO$_3$/WN high-k/metal gate stacks were all deposited by atomic layer deposition (ALD) in a tubular reactor without air break between layers. A short-time buffered oxide etch and 10% (NH$_4$)$_2$S passivation were performed before transferring the samples to the ALD reactor. The Al$_2$O$_3$ and LaAlO$_3$ were grown at 300 $^o$C and the WN was grown at 385$^o$C. Lanthanum tris(*N,N'*-diisopropylformamidinate), trimethylaluminum, and H$_2$O were used as the precursors for LaAlO$_3$ deposition, while bis(*tert*-butylimido)bis(dimethylamido)tungsten(VI) vapor and ammonia gas were use as the WN precursors. The sheet resistance of the ALD WN film was measured by a four-point probe station, and the resistivity was 2.2 mΩ·cm for a 40 nm WN film. The excellent conformity and uniformity of the ALD process is critical for realizing the GAA structure with ultra-small EOT. The $L_{ch}$ is varied from 80nm down to 20nm, $W_{NW}$ is varied from 35nm down to 20nm with a fixed nanowire height ($H_{NW}$) of 30nm defined by the MBE thickness. The nanowire length ($L_{NW}$) is fixed at 200nm. Four parallel wires are integrated in each device. The nanowires are aligned along [100] direction as required by the anisotropic release process, which also defines the transport direction of the GAA devices. Fig. 3 shows the top-view SEM images of the 20nm PMMA mask defining the smallest $L_{ch}$ and a finished GAA device. Note that due to the dopant diffusion, the actual $L_{ch}$ is smaller than the defined $L_{ch}$. All patterns were defined by a Vistec VB6 UHR electron beam lithography system and dry etching was performed with a Panasonic E620 high density plasma etcher. The MOSFET electrical characterization was performed using a Keithley 4200 at room temperature.

## Results and Discussion

Fig. 4-6 show the well-behaved output characteristics, transfer characteristics, and $g_m$-$V_{gs}$ of a GAA FET (sample A) with $L_{ch}$ = $W_{NW}$ = 20nm. The current is normalized by the total perimeter of the nanowires, i.e. $W_G = 2\times(W_{NW} + H_{NW})\times$(Wire Number). The 20nm $L_{ch}$ device shows negligible channel length modulation, $I_{ON} = 850\mu A/\mu m$ at $V_{DD} = 0.8V$, $g_m = 1.65mS/\mu m$ at $V_{ds} = 0.5V$, SS = 75mV/dec and DIBL = 40mV/V. The device operates in enhancement-mode with $V_T = 0.14V$ extracted by linear extrapolation at $V_{ds} = 0.05V$. Fig. 7 shows the SS scaling metrics for sample A with $W_{NW} = 20nm$. Error bars show the standard deviation of the measurement over 25 devices at each data point. Consistent sub-100mV/dec SS has been obtained for the devices at $V_{ds}=0.5V$ for all $L_{ch}$. Figs. 8 - 9 show the SS and DIBL scaling metrics with different $W_{NW}$. No evident $W_{NW}$ dependence is observed, indicating the current EOT and GAA structure yield a very small geometric screening length ($\lambda$) compared to $L_{ch}$. The mean SS and DIBL remain unchanged at ~75mV/dec and ~25mV/V with $L_{ch}$ down to 50nm, showing the immunity of these devices to SCE.

Fig. 10 shows the transfer characteristics of one of the three best devices with SS of 63mV/dec at $V_{ds}=0.05V$, indicating excellent gate control and low $D_{it}$. Seventeen devices show SS below 70mV/dec. The lowest DIBL achieved is 7mV/V (not shown). Figs. 11 - 12 show that $g_m$ and $I_{ON}$ remain constant at small $L_{ch}$ indicating near-ballistic transport. Fig. 13 shows the $D_{it}$ box plot and histogram of sample A. The midgap $D_{it}$ is extracted from the SS at $V_{ds} = 0.05V$ for $L_{ch} = 50 - 80nm$ since these devices are immune to SCE. A mean $D_{it}$ of ~$4\times10^{12}$ $eV^{-1}cm^{-2}$ is obtained with the lowest value of $9\times10^{11}eV^{-1}cm^{-2}$ corresponding to the 63mV/dec device. Fig. 14 shows the increasing $g_m$ and $I_{on}$ with decreasing $W_{NW}$, due to the quantum confinement and volume inversion effect [2]. Simulation of inversion charge distribution inside the InGaAs nanowires in Fig. 15 (a) confirms volume inversion at all $W_{NW}$. Fig. 15 (b) shows the inversion layer distribution along $y = 15nm$ for different $W_{NW}$. It is projected that the optimum $W_{NW}$ occurs at around 10nm where a plateau across the entire nanowire would form. Further reduction of nanowire size is therefore required to study the ultimate performance limit of InGaAs GAA devices.

Figs. 16 - 17 show the transfer characteristics and $g_m$-$V_{gs}$ for 20nm $L_{ch}$ InGaAs GAA MOSFETs with EOT = 1.7nm (sample C). Increasing EOT has led to increased SCE, evident from the larger SS and DIBL. However, higher $g_m = 1.8mS/\mu m$ at $V_{ds} = 0.5V$ and $2.1mS/\mu m$ at $V_{ds} = 1V$ is obtained on sample C, indicating enhanced mobility with relaxed EOT. A negative $V_T = -0.138V$ is also obtained from the same device. Fig. 18 shows the relatively low gate leakage current density for EOT=1.2 and 1.7nm even with the advanced 3D structure. Figs. 19 - 21 show the SS, DIBL, and $V_T$ scaling metrics for sample A, B, and C with $L_{ch}$ ranging from 20 to 80nm. Sample C shows the largest SS and DIBL due to the larger EOT of 1.7nm. Sample B shows worse SS and DIBL compared to sample A with the same EOT = 1.2nm due to the larger $D_{it}$ [3]. This indicates that insertion of ultra-thin $Al_2O_3$ interfacial layer can effectively improve LaAlO$_3$/InGaAs interface quality. Decreasing EOT also results in better $V_T$ roll-off properties and is favorable for an enhancement-mode operation. It is also noted that the variation of sample A is also the least among all three gate stacks. This indicates that the EOT scaling with effective interface passivation has led to not only scalability improvements but also a variability breakthrough. Detailed analysis of the variability and reliability of InGaAs nanowire devices is on-going. Fig. 22 benchmarks SS and DIBL in this work with InGaAs non-planar FETs fabricated in our group. The GAA structure with thin EOT has shown significant improvement in the control of SCE, due to the better gate control. Table 1 compares the device dimensions and performance in this work with the representative non-planar and thin body planar InGaAs MOSFETs in the literature [4-8]. The successful demonstration of the smallest $L_{ch}$, $W_{NW}$, SS, DIBL and highest $g_m$ has been achieved in this work.

## Conclusion

We have demonstrated the shortest $L_{ch}$=20nm InGaAs GAA nanowire MOSFETs with ALD $Al_2O_3$/LaAlO$_3$ gate stack. Lowest SS of 63mV/dec and DIBL of 7mV/V have been achieved. Benefiting from both the ultimate scalability of GAA structure and excellent transport property of III-V channel, InGaAs GAA technology is a strong candidate for future high-speed low-power logic applications.

## Acknowledgement

The authors would like to thank Y. Q. Wu, X. L. Li, M. S. Lundstrom, D. A. Antoniadis, and J. A. del Alamo for the valuable discussions. This work is supported by the SRC FCRP MSD Center, NSF and AFOSR.

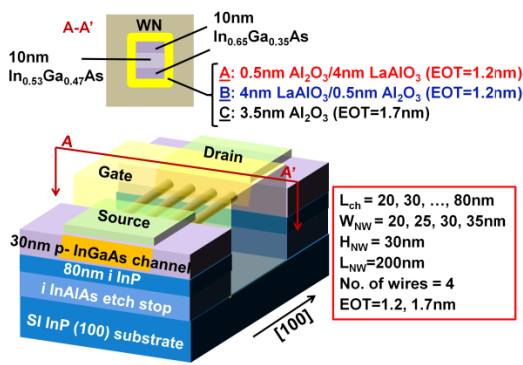

Fig. 1 Device structure, dimension and key parameters of InGaAs GAA MOSFETs.

Fig. 2 Fabrication process flow and device splits of InGaAs GAA MOSFETs.

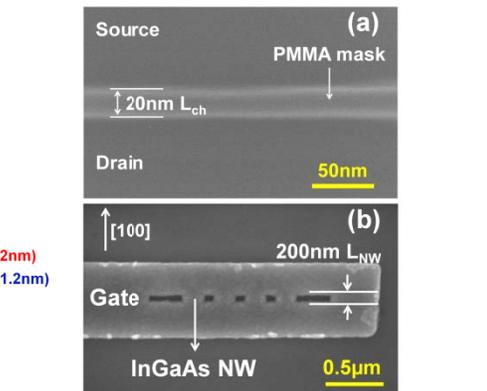

Fig. 3 SEM images of (a) PMMA mask defining 20nm $L_{ch}$ (b) an InGaAs GAA FET with 4 wires.

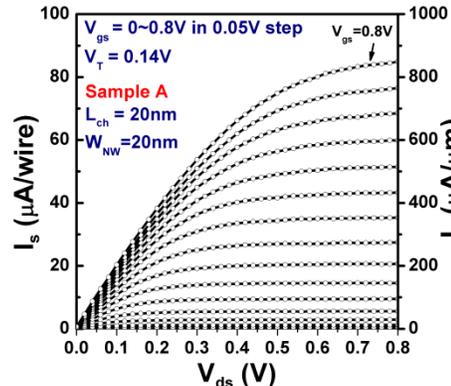

Fig. 4 Output characteristics of a 20nm $L_{ch}$ InGaAs GAA MOSFET with $Al_2O_3/LaAlO_3$ gate dielectric (Sample A, EOT=1.2nm) and $W_{NW}$=20nm. $I_s$ is used due to relatively large junction leakage current in $I_d$.

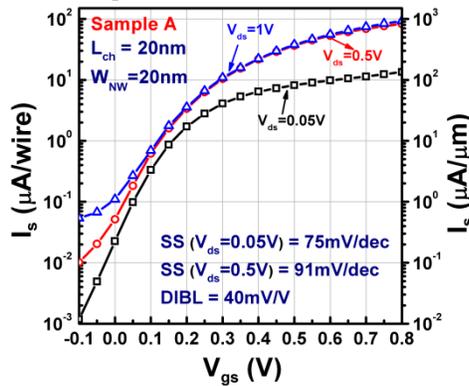

Fig. 5 Transfer characteristics of the same device shown in Fig. 4. $W_G$=100nm for $W_{NW}$=20nm normalized to perimeter.

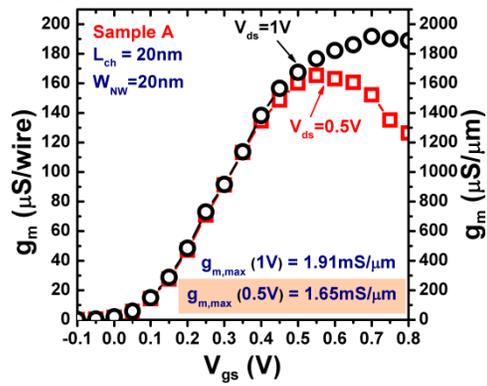

Fig. 6 Transconductance of the same device shown in Fig. 4.

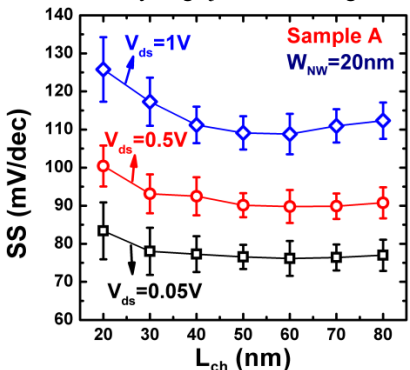

Fig. 7 SS scaling metrics for InGaAs GAA MOSFETs (Sample A, $W_{NW}$=20nm). Error bars show standard deviation of the measurements over 25 devices.

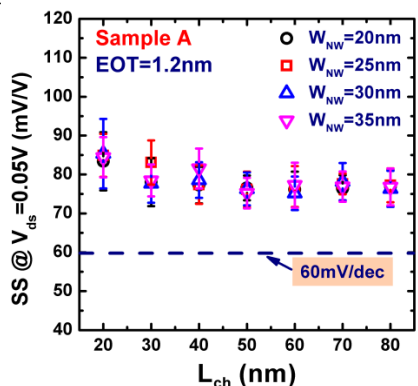

Fig. 8 SS ($V_{ds}$=0.05V) scaling metrics for sample A with $W_{NW}$ from 20 to 35nm.

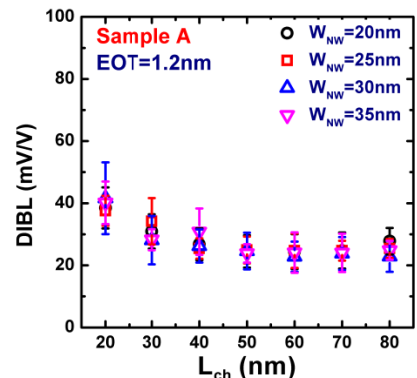

Fig. 9 DIBL scaling metrics for sample A with $W_{NW}$ from 20 to 35nm.

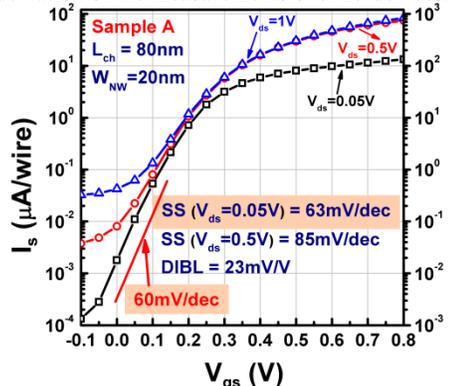

Fig. 10 Transfer characteristics of $L_{ch}$=80nm InGaAs GAA MOSFET (Sample A, $W_{NW}$=20nm) with SS=63mV/dec at $V_{ds}$=0.05V. Three measured devices show SS=63mV/dec.

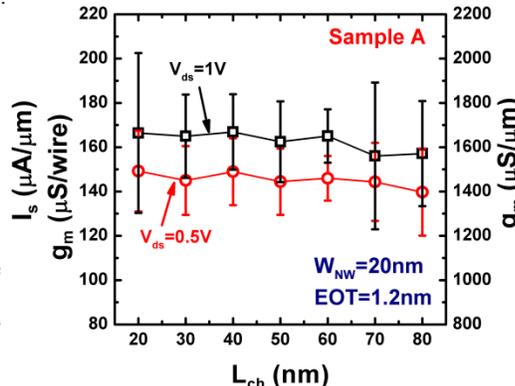

Fig. 11 $g_m$ scaling metrics for sample A (EOT=1.2nm, $W_{NW}$=20nm) at $V_{ds}$=0.5V and 1V.

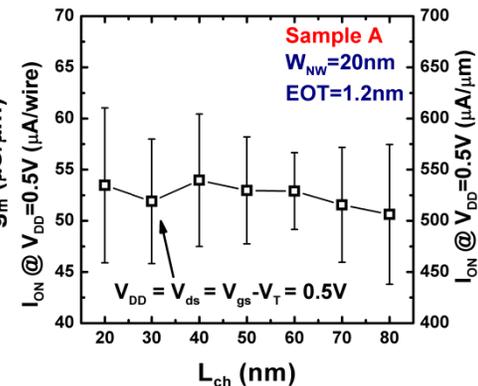

Fig. 12 $I_{ON}$ scaling metrics for sample A (EOT=1.2nm, $W_{NW}$=20nm) at $V_{ds}=V_{gs}-V_T$=0.5V.

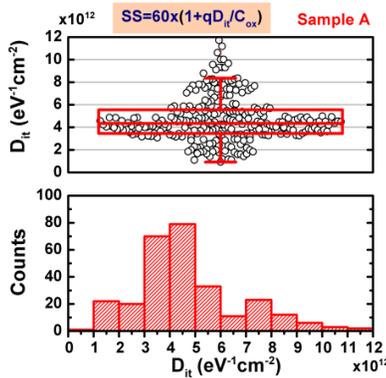

Fig. 13 Box plot and histogram for mid-gap $D_{it}$ of sample A ($W_{NW}$=20-35nm, $L_{ch}$=50-80nm) extracted from SS. 3 devices have SS=63 mV/dec and 17 devices have SS<70 mV/dec.

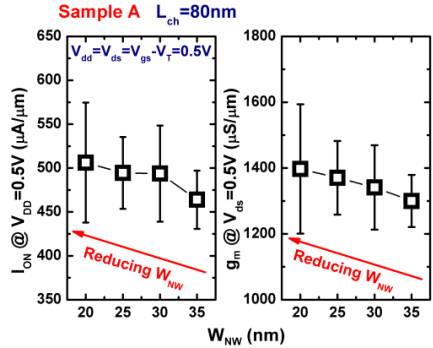

Fig. 14 $W_{NW}$ dependence of the $I_{ON}$ ($V_{DD}$=0.5V) and $g_m$ ($V_{ds}$=0.5V), showing improved performance when reducing nanowire size.

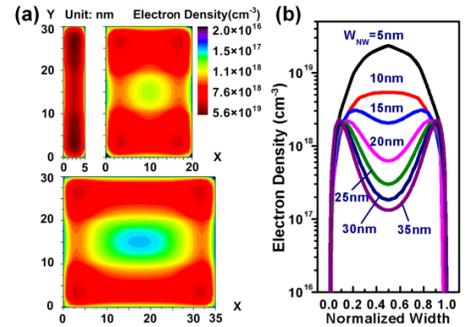

Fig. 15 (a) Simulated inversion charge distribution for current InGaAs nanowire MOSFETs with $W_{NW}$=5, 20, and 35nm, indicating volume inversion for all cases (b) Electron distribution along y=15nm for $W_{NW}$=5-35nm in 5nm step.

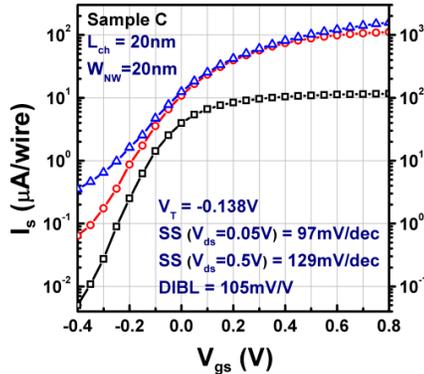

Fig. 16 Transfer characteristics of 20nm $L_{ch}$ InGaAs GAA MOSFET with $Al_2O_3$ gate (Sample C, EOT=1.7nm) and $W_{NW}$=20nm.

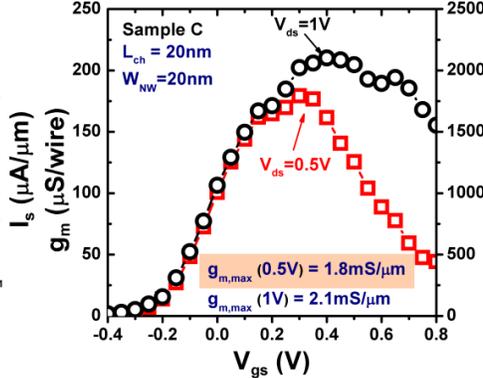

Fig. 17 Transconductance of 20nm $L_{ch}$ InGaAs GAA MOSFET with $Al_2O_3$ gate (Sample C, EOT=1.7nm) and $W_{NW}$=20nm.

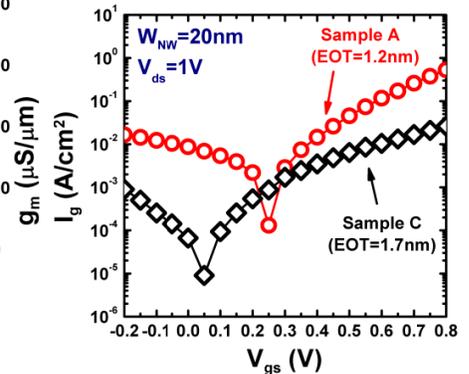

Fig. 18 Gate leakage of sample A (EOT=1.2nm) and sample C (EOT=1.7nm).

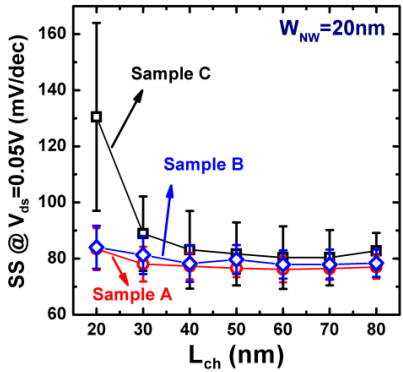

Fig. 19 SS scaling metrics for samples A, B, and C with $W_{NW}$=20nm.

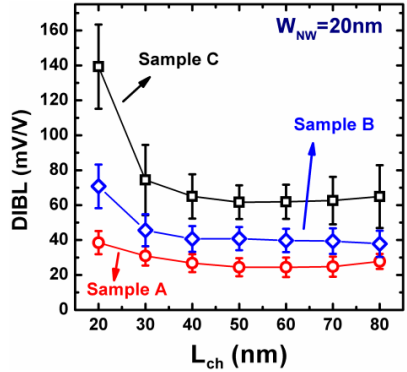

Fig. 20 DIBL scaling metrics for sample A, B, and C with $W_{NW}$=20nm.

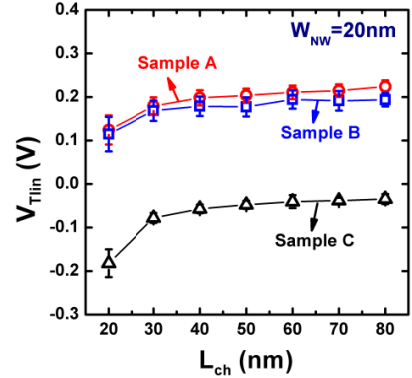

Fig. 21 $V_T$ scaling metrics for sample A, B, and C with $W_{NW}$=20nm.

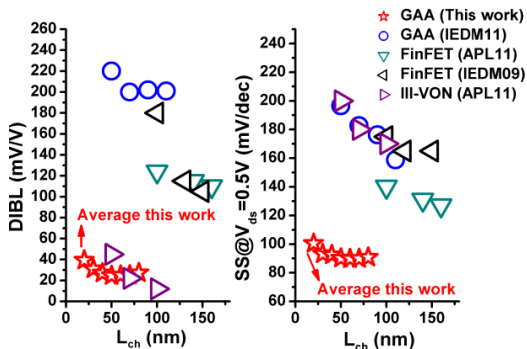

Fig. 22 SS/DIBL-$L_{ch}$ benchmarking for InGaAs GAA nanowire MOSFETs in this work with other 3D InGaAs MOSFETs demonstrated at Purdue. Much lower SS, DIBL is obtained at smaller $L_{ch}$ due to the GAA structure and thin EOT.

Table 1: Performance benchmark of typical non-planar and ETB InGaAs MOSFETs

|  | This work** | Ref. [1] | Ref. [4] | Ref. [5] | Ref. [6] | Ref. [7] | Ref. [8] |
|---|---|---|---|---|---|---|---|
| $In_xGa_{1-x}As$ (x) | 0.65 | 0.53 | 0.7 | 0.7 | 1 | 0.7 | 1 |
| Structure | GAA | GAA | Tri-gate | GAA | GAA | FinFET | ETB |
| Fabrication | Top-down | Top-down | Top-down | Bottom-up | Bottom-up | Top-down | Top-down |
| $L_{ch,min}$ (nm) | 20 | 50 | 60 | 200 | 100 | 130 | 55 |
| $W_{NW(Fin),min}$ (nm) | 20 | 30 | 30 | 90 | 15 | 220 | - |
| EOT (nm) | 1.2 | 4.5 | 1.2 | 1.8* | 1.1* | 4.5* | 3.5 |
| SS [$V_{ds}$=0.5V] (mV/dec) | 88 | 245 | 94* | 98 | 140 | - | - |
| SS [$V_{ds}$=0.05V] (mV/dec) | 63 | 145 | 66 | 90* | - | 230* | 105 |
| DIBL (mV/V) | 7 | 210 | 60* | 170* | 60 | 135 | 84 |
| $g_{m,max}$ (mS/μm) [$V_{ds}$=0.5V] | 1.74 | 0.45 | - | - | 1.23 | - | - |

*Extracted/estimated from literature
**Reported values are best from all measured devices